\documentstyle[12pt,amssymbols]{article}
\newcommand{\bq}{\begin{equation}}
\newcommand{\eq}{\end{equation}}
\newcommand{\bqa}{\begin{eqnarray}}
\newcommand{\eqa}{\end{eqnarray}}
\newcommand{\ra}{\rightarrow}

\def\g{\gamma}

\def\ov{\over}

\def\ed{\end{document}}

\def\ra{\rightarrow}

\def\2pi{1\over 2\pi i}
\def\q{q-q^{-1}}
\def\newline{\hfil\break}

\def\ra{\rightarrow}
\def\va{\varphi}

\def\sq2{\sqrt{2}}
\def\sqk2{\sqrt{2(k+2}}
\def\sqk{\sqrt{k}}

\def\be{\begin{equation}}
\def\ee{\end{equation}}
\def\br{\begin{array}}
\def\er{\end{array}}
\def\bea{\begin{eqnarray}}
\def\eea{\end{eqnarray}}
\def\ba{\begin{equation}\begin{array}}
\def\ea{\end{array}\end{equation}}
\def\bac{\begin{equation}\begin{array}{rll}}

\newcommand{\LS}{\widehat{sl(2)}}
\newcommand{\uq}{U_q (\widehat{sl(2)})}

\def\Z{{\Bbb Z}}

\begin{document}
\begin{titlepage}
%\rightline{DRAFT}
\rightline{hep-th/}
\rightline{CRM-2201}
\rightline{July 20, 1994}
\vbox{\vspace{-10mm}}
\vspace{1.0truecm}
\begin{center}
{\LARGE \bf ON A BOSONIC-PARAFERMIONIC REALIZATION OF $\uq$ }\\[8mm]
{\large A. Hamid Bougourzi$^1$  and  Luc Vinet$^2$}\\
[6mm]{\it Centre de Recherches Math\'ematiques,
Universit\'e de Montr\'eal\\
C.P. 6128-A, Montr\'eal (Qu\'ebec) H3C 3J7, Canada.}\\[20mm]
\end{center}
\vspace{1.0truecm}
\begin{abstract}
We realize the $U_q(\widehat{sl(2)})$ current algebra at arbitrary
level in terms of one deformed free bosonic field and a pair of
deformed parafermionic fields. It is shown that the
operator product expansions of these
parafermionic fields involve
an infinite number of simple poles and simple zeros, which then condensate
to form a branch cut in the classical limit $q\rightarrow 1$. Our realization
coincides with those of Frenkel-Jing and Bernard when the level $k$
takes the values 1 and 2 respectively.
\end{abstract}

%%%%%%%%%%%%%%%%%%%%%%%%%%%%%%%%%%%%%%%%%%%%%%%%%%%%%%%
\footnotetext[1]{
Email: {\tt bougourz@ere.umontreal.ca}
}
\footnotetext[2]{
Email: {\tt vinet@ere.umontreal.ca}
}
\end{titlepage}
\section{Introduction}

The free field realization of quantum affine algebras
\cite{Dri85,Jim85,Dri86}
has attracted a lot of attention in recent years due to
both its physical and mathematical applications.
Most of the realizations are given in terms of free
bosonic fields and are applied either to simply laced
quantum affine algebras at level 1, or to $\uq$ and more recently
$U_q(\widehat{sl(n)})$ at arbitrary level
\cite{FrJi88,Abaal92,Mat92,AOSal93}. (See also Ref. \cite{Bou93}
for more references and the equivalence of all
the bosonic realizations.) The extension of
such bosonic realizations to other quantum affine algebras
has not yet been achieved and it is believed to be highly
complicated.

It is therefore
natural to investigate the realizations of quantum
affine algebras either in terms of other types of fields such as
fermionic, ghost and parafermionic ones, or in
terms of a combination of the latter fields and bosonic ones.
These realizations might
be simpler than the pure bosonic ones and also applicable to
more general quantum affine algebras. In fact, a
fermionic and ghost realization of
$U_q(\widehat{sl(n)}$ at level zero is achieved
in Ref. \cite{Hay90}, a mixed bosonic-fermionic realization of
$U_q(so\widehat{(2n+1)})$ at level 1 is derived in
Ref. \cite{Ber89}, and more recently a pure fermionic realization
of $\uq$ at level 2 has been provided in Ref. \cite{BoVi94}.

We pursued these investigations and present here a
bosonic-parafermionic realization of $\uq$ at arbitrary level.
It is known that in the classical case, in addition to $\uq$
at arbitrary level, any untwisted non-simply laced affine algebra
at level 1 and any twisted simply laced affine algebra at level 2
can be realized with the help of both bosonic and parafermionic fields
\cite{BeTh87,Godal87}. As will be seen in this paper, a model for
the simplest quantum affine algebra ($\uq$)  can be naturally devised
by introducing a deformed parafermionic field. We trust that the
results reported in this paper will be of use in the construction
of  realizations of quantum deformations of the aforementioned
affine algebras.

We describe in section 2
the $\uq$ quantum affine algebra at arbitrary level.
We then derive in section 3
a  bosonic-parafermionic realization of this
algebra which is the
quantum analogue of the classical bosonic-parafermionic
realization of $\LS$ given in Ref. \cite{ZaFa85}.

\section{The $\uq$ quantum current algebra}

The $\uq$ affine algebra  is the unital
associative algebra
generated
by the  elements
$\{ E^{\pm}_n,H_m, K^{\pm},
q^{\pm d}, \g^{\pm 1/2}; n\in \Z,~m \in
\Z_{ \neq 0}\}$ that obey the defining relations
\cite{Dri88}
\bac
&&{[H_n,H_m]} = {[2n]\over 2n} {{\g^{n}-\g^{-n}}
\ov {q-q^{-1}}}
\delta_{n+m,0},\\
&&KH_nK^{-1}=  H_n,\\
&& {[H_n,E^{\pm}_m]}=
\pm\sqrt{2}{\g^{\mp |n|/2}[2n]\over 2n}
E^\pm_{n+m},\\
&& KE^\pm_n K^{-1} =q^{\pm 2} E^\pm_n,\\
&&{[E^+_n,E^-_m]} = {\g^{(n-m)/2}\Psi_{n+m}-\g^{(m-n)/2}
\Phi_{n+m}\over q-q^{-1}},\\
&&E^\pm_{n+1}E^\pm_m-q^{\pm 2}E^\pm_mE^\pm_{n+1}=
q^{\pm 2}E^\pm
_nE^\pm_{m+1}-E^\pm_{m+1}E^\pm_n,\\
&&q^dE^\pm_nq^{-d}=q^nE^\pm_n,\\
&&q^dH_nq^{-d}=q^nH_n,\\
&&{[\g^{\pm 1/2}, x]}=0,\quad \forall x\in \uq,
\label{cwb}
\ea
with
$[n]\equiv (q^n-q^{-n})/(q-q^{-1})$
and where $\Psi_n$ and $\Phi_n$
are given by
the mode expansions of the fields
$\Psi(z)$ and $\Phi(z)$, which are themselves defined by
\bac
\Psi(z)&=&\sum\limits_{n\geq 0}\Psi_nz^{-n}=K
\exp\{\sqrt{2}(\q)\sum\limits_{n>0}H_nz^{-n}\},\\
\Phi(z)&=&\sum\limits_{n\leq 0}\Phi_nz^{-n}=K^{-1}
\exp\{-\sqrt{2}(\q)\sum\limits_{n<0}H_nz^{-n}\}.
\label{algebra}
\ea
The central element $\gamma$ acts as the scalar $q^k$ on the
level $k$ representations of $\uq$.
%This algebra can be re-expressed as a quantum current algebra, which is
% most convenient
For the purpose of constructing field realizations, it is most
convenient to re-express this algebra as a quantum current algebra and
to define it
 through the following operator product expansions
(OPE's):
\bac
\Psi(z).\Phi(w)&=&
{(z-wq^{2+k})(z-wq^{-2-k})\over (z-wq^{2-k})(z-wq^{-2+k})}
\Phi(w).\Psi(z), \\
\Psi(z).E^{\pm}(w)&=&
q^{\pm 2}{(z-wq^{\mp(2+k/2)})\over z-wq^{\pm (2-k/2)}}
E^\pm(w).\Psi(z), \\
\Phi(z).E^{\pm}(w)&=&
q^{\pm 2}{(z-wq^{\mp(2-k/2)})\over z-wq^{\pm (2+k/2)}}
E^\pm(w).\Phi(z),\\
E^+(z).E^-(w)&\sim& {1
\over w(\q)}\left\{{\Psi(wq^{k/2})\over z-wq^k}-
{\Phi(wq^{-k/2})\over z-wq^{-k}}\right\},\quad |z|>|wq^{\pm k}|,
\\
E^-(z).E^+(w)&\sim& {1
\over w(\q)}\left\{{\Psi(wq^{k/2})\over z-wq^k}-
{\Phi(wq^{-k/2})\over z-wq^{-k}}\right\},\quad |z|>|wq^{\pm k}|,
\\
E^{\pm}(z).E^{\pm}(w)&=&{(z q^{\pm 2}-w)\over z-w q^{\pm 2}}
E^{\pm}(w). E^{\pm}(z).
\label{op6}
\ea
Here, the quantum currents $E^\pm(z)$ are the
following generating functions of the  generators $E^\pm_n$:
\be
E^\pm(z)=\sum_{n\in\Z}E^\pm_nz^{-n-1},
\ee

\subsection{A bosonic-parafermionic realization of $\uq$}

Let $\va^{\pm}(z)$ denote two different deformations of the
same bosonic free field and take their mode expansions to be
given by
\be
\va^{\pm}(z)=\va-i\va_0\ln{z}
+ik\sum_{n\neq 0}{q^{\mp |n|k/2}\over [nk]}\va_nz^{-n}.
\ee
The operators $\{\va, \va_n;\> n\in \Z\}$
generate a deformed Heisenberg  algebra with
the following commutation relations:
\bac
{[\va_n,\va_m]}&=&{[2n][nk]\over 2kn}\delta_{n+m,0}\>,\\
{[\va,\va_0]}&=&i\>.
\label{qhma}
\ea
 Let
\be
V^\pm(a,z)\equiv :e^{ia\va^{\pm}(z)}:
\ee
denote a normal ordered vertex operator with
$a$  an arbitrary real number.
The bosonic normal ordering  symbol
:: indicates that for fields between the two colons,
 the creation modes $\{\va_n, \va;n<0\}$ should
be moved to the left of the annihilation modes
$\{\va_n; n\geq 0\}$.
Henceforth, we will  use the standard convention that
operators defined at the same point are understood
to be normal ordered. Using this definition of normal ordering,
one can derive the following OPE of vertex operators of the
exponential type:
\bac
V^{\pm}(a,z).V^{\mp}(b,w)&=&
e^{-ab<\va^{\pm}(z)\va^{\mp}(w)>}
:V^{\pm}(a,z)V^{\mp}(b,w):,\\
V^{\pm}(a,z).V^{\pm}(b,w)&=&
e^{-ab<\va^{\pm}(z)\va^{\pm}(w)>}
:V^{\pm}(a,z)V^{\pm}(b,w):,
\ea
with the vacuum expectation values  given by
\bac
<\va^{\pm}(z)\va^{\mp}(w)>&=&
\va^{\pm}(z).\va^{\mp}(w)-:\va^{\pm}(z)\va^{\mp}(w):\\
&=&-\ln{z}+{k\over 2}\sum_{n>0}{[2n]\over n[nk]}w^nz^{-n}\\
&=&-\ln{z}+{k\over 2}\sum_{n\geq 0}ln\left({1-q^{k+2+2nk}wz^{-1}
\over 1-q^{k-2+2nk}wz^{-1}}\right),\\
<\va^{\pm}(z)\va^{\pm}(w)>&=&
\va^{\pm}(z).\va^{\pm}(w)-:\va^{\pm}(z)\va^{\pm}(w):\\
&=&-\ln{z}+{k\over 2}\sum_{n>0}{q^{\mp nk}[2n]\over n[nk]}w^nz^{-n}\\
&=&-\ln{z}+{k\over 2}\sum_{n\geq 0}ln\left({1-q^{2+2nk+k\mp k}wz^{-1}
\over 1-q^{-2+2nk+k\mp k}wz^{-1}}\right).
\ea
Here, without loss of generality, we have assumed that
$|q|<1$.
Our bosonic-parafermionic realization of $\uq$ at
arbitrary level is
given by
\bac
\!\!\!\!\Psi(z)&= &:V^+(\sqrt{2\over k},zq^{k/2})
V^-(-\sqrt{2\over k},zq^{-k/2}):\\
&=&
q^{\sqrt{2k}\va_0}\exp\left(
\sqrt{2k}(\q)\sum_{n>0}\va_nz^{-n}\right),\\
\!\!\!\!\Phi(z)&= &
:V^+(\sqrt{2\over k},zq^{-k/2})
V^-(-\sqrt{2\over k},zq^{k/2}):\\
&=&q^{-\sqrt{2k}\va_0}
\exp\left(-\sqrt{2k}(\q)\sum_{n<0}\va_nz^{-n}\right),\\
\!\!\!\!E^{\pm}(z)&=&\sqrt{[k]}\psi^{\pm}(z)
V^{\pm}(\pm\sqrt{2\over k},z),
\label{ABE}
\ea
where the `basic' fields $V^{\pm}(\pm \sqrt{2\over k},
z)$ and $\psi^{\pm}(z)$ satisfy the following OPE's:
\bac
V^{\pm}(\pm\sqrt{2\over k},z).V^{\mp}(\mp
\sqrt{2\over k},w)&=&
z^{-{2\over k}}{(q^{k+2}wz^{-1};q^{2k})_{\infty}\over
(q^{k-2}wz^{-1};q^{2k})_{\infty}}
:V^{\pm}(\pm\sqrt{2\over k},z)V^{\mp}(\mp
\sqrt{2\over k},w):,\\
V^{\pm}(\pm\sqrt{2\over k},z).V^{\pm}(\pm
\sqrt{2\over k},w)&=&
z^{{2\over k}}{(q^{k\mp k-2}wz^{-1};q^{2k})_{\infty}\over
(q^{k\mp k+2}wz^{-1};q^{2k})_{\infty}}
:V^{\pm}(\pm\sqrt{2\over k},z)V^{\pm}(\pm
\sqrt{2\over k},w):,\\
\psi^{\pm}(z).\psi^{\mp}(w)&=&
z^{2\over k}
{(q^{k-2}wz^{-1};q^{2k})_{\infty}\over (q^{k+2}wz^{-1};q^{2k})_
{\infty}}\left({1\over (z-wq^{k})(z-wq^{-k})}
+{\rm regular}\right) ,\\
\psi^{\pm}(z).\psi^{\pm}(w)&=&
{(wz^{-1})^{2\over k}(zq^{\pm 2}-w)\over
z-wq^{\pm 2}}
{(q^{k\mp k-2}zw^{-1};q^{2k})_{\infty}
(q^{k\mp k+2}wz^{-1};q^{2k})_{\infty}\over
(q^{k\mp k+2}zw^{-1};q^{2k})_{\infty}
(q^{k\mp k-2}wz^{-1};q^{2k})_{\infty}}
\psi^{\pm}(w).\psi^{\pm}(z),\\
V^{\pm}(\pm\sqrt{2\over k},z).\psi^{\pm}(w)
&=&\psi^{\pm}(w).V^{\pm}(\pm\sqrt{2\over k},z)={\rm regular},\\
V^{\pm}(\pm\sqrt{2\over k},z).\psi^{\mp}(w)
&=&\psi^{\mp}(w).V^{\pm}(\pm\sqrt{2\over k},z)={\rm regular},
\label{OPEP}
\ea
and where $(a,q)_{\infty}$ stands for the infinite product
\be
(a,q)_{\infty}\equiv \prod_{n=0}^{\infty}(1-aq^n),\quad\quad |q|<1\> .
\ee
It can easily be checked that with these OPE's
the quantum currents $E^\pm(z)$, $\Psi(z)$ and
$\Phi(z)$ as defined by (\ref{ABE}) do indeed
satisfy the $\uq$ quantum current algebra
(\ref{op6}). Moreover, let us
further verify that the above realization reduces to that
of Frenkel and Jing when $k=1$ \cite{FrJi88}, and to that
of Bernard when $k=2$ \cite{Ber89}. The OPE's
given by (\ref{OPEP}) take the following
simple forms when $k=1$:
\bac
V^{\pm}(\pm\sqrt{2},z).V^{\mp}(\mp
\sqrt{2},w)&=&
{1\over (z-wq)(z-wq^{-1})}
:V^{\pm}(\pm\sqrt{2},z)V^{\mp}(\mp
\sqrt{2},w):,\\
V^{\pm}(\pm\sqrt{2},z).V^{\pm}(\pm
\sqrt{2},w)&=&
(z-wq^{1\mp 1})(z-wq^{-1\mp 1})
:V^{\pm}(\pm\sqrt{2},z)V^{\pm}(\pm
\sqrt{2},w):,\\
\psi^{\pm}(z).\psi^{\mp}(w)&=&
1+(z-wq)(z-wq^{-1})(\rm regular),\\
\psi^{\pm}(z).\psi^{\pm}(w)&=&
\psi^{\pm}(w).\psi^{\pm}(z),\\
V^{\pm}(\pm\sqrt{2},z).\psi^{\pm}(w)
&=&\psi^{\pm}(w).V^{\pm}(\pm\sqrt{2},z)={\rm regular},\\
V^{\pm}(\pm\sqrt{2},z).\psi^{\mp}(w)
&=&\psi^{\mp}(w).V^{\pm}(\pm\sqrt{2},z)={\rm regular}.
\ea
It is  clear from these  relations
that the parafermionic fields $\psi^{\pm}(z)$ can here be
identified with the identity operator.
The Frenkel-Jing realization
\bac
\!\!\!\!\Psi(z)&= &:V^+(\sqrt{2},zq^{1/2})
V^-(-\sqrt{2},zq^{-1/2}):\\
&=&
q^{\sqrt{2}\va_0}\exp\left(
\sqrt{2}(\q)\sum_{n>0}\va_nz^{-n}\right),\\
\!\!\!\!\Phi(z)&= &
:V^+(\sqrt{2},zq^{-1/2})
V^-(-\sqrt{2},zq^{1/2}):\\
&=&q^{-\sqrt{2}\va_0}
\exp\left(-\sqrt{2}(\q)\sum_{n<0}\va_nz^{-n}\right),\\
\!\!\!\!E^{\pm}(z)&=&
V^{\pm}(\pm\sqrt{2},z).
\ea
is thus recovered in this case.
When $k=2$, the OPE's (\ref{OPEP}) also reduce to
 the following
simple expressions:
\bac
V^{\pm}(\pm 1,z).V^{\mp}(\mp 1,w)&=&
{1\over (z-w)}
:V^{\pm}(\pm 1,z)V^{\mp}(\mp
1,w):,\\
V^{\pm}(\pm 1,z).V^{\pm}(\pm
1,w)&=&
(z-wq^{\mp 2})
:V^{\pm}(\pm 1,z)V^{\pm}(\pm
1,w):,\\
\psi^{\pm}(z).\psi^{\mp}(w)&=&
{z-w\over (z-wq^2)(z-wq^{-2})}
+(z-w)(\rm regular),\\
\psi^{\pm}(z).\psi^{\pm}(w)&=&
-\psi^{\pm}(w).\psi^{\pm}(z),\\
V^{\pm}(\pm 1,z).\psi^{\pm}(w)
&=&\psi^{\pm}(w).V^{\pm}(\pm 1,z)={\rm regular},\\
V^{\pm}(\pm 1,z).\psi^{\mp}(w)
&=&\psi^{\mp}(w).V^{\pm}(\pm 1,z)={\rm regular}.
\ea
These  relations now suggest  that it is
possible to make the following identifications:
\be
\psi^{\pm}(z)\equiv\psi(z),
\ee
where $\psi(z)$ is a deformed real free fermionic
field such that
\be
\psi(z).\psi(w)=
{z-w\over (z-wq^2)(z-wq^{-2})}
+:\psi(z)\psi(w):,
\ee
with $:\psi(z)\psi(z):=0$. We are here using the (same)
symbol :: to denote the fermionic normal ordering.
If
\be
\psi(z)=\sum_{r\in \Z+1/2}\psi_r z^{-r-1/2}\>,
\ee
this
normal ordering is defined
so as to have the fermionic creation modes
 $\{\psi_r, r\leq -1/2\}$
of fields between the two colons put to the left of their annihilation modes
$\{\psi_r, r\geq 1/2\}$.
Up to a trivial scaling factor, this
deformed fermionic field is the same as the one introduced by
Bernard \cite{Ber89} in his mixed bosonic-fermionic
realization of $U_q(so
\widehat{(2n+1)})$ at level 1,
and in particular in the boson-fermion realization of
$\uq$ at level 2, which reads
\bac
\!\!\!\!\Psi(z)&= &:V^+(1,zq)
V^-(-1,zq^{-1}):\\
&=&
q^{2\va_0}\exp\left(
2(\q)\sum_{n>0}\va_nz^{-n}\right),\\
\!\!\!\!\Phi(z)&= &
:V^+(1,zq^{-1})
V^-(-1,zq):\\
&=&q^{-2\va_0}
\exp\left(-2(\q)\sum_{n<0}\va_nz^{-n}\right),\\
\!\!\!\!E^{\pm}(z)&=&\sqrt{[2]}\psi(z)
V^{\pm}(\pm 1,z).
\ea

Note that in the limit
$q\ra 1$, the OPE's of $\psi^\pm(z)$ and $\psi^\mp(w)$ given
in (\ref{OPEP}) take
the form
\be
\psi^\pm(z).\psi^\mp(w)=(z-w)^{2\over k}({1\over
(z-w)^2}+{\rm regular}).
\label{pif}
\ee
In the quantum case, an infinite number of alternating simple poles and
of simple zeros occur in the expansions of $\psi^\pm(z).\psi^\mp(w)$.
They are found respectively at $z=wq^{k+2+2kn};\>n\geq 0$ and
$z=wq^{k-2+2kn};\>n\geq 0$, and when $q$ is real, they are seen to
lay on the segment connecting $w$ and the origin. It follows from
(\ref{pif}), that these zeros and poles condensate to form a branch cut
in the classical limit $q\ra 1$.

It would be interesting to extend this
bosonic-parafermionic realization of $\uq$
to other quantum affine algebras as discussed in the introduction.
One might also look at the relation between this realization of
$\uq$
and the  bosonic ones.
They could be equivalent as in the classical situation (see \cite{Bil89}),
owing to the existence of a
  boson-parafermion correspondence.
Let us finally mention that in the classical case,
the full parafermionic algebra is generated (see \cite{ZaFa85,Bil89})
by a set of fields $\psi^{\pm}_n(z)$,
$0\leq n\leq k$, with $\psi^{+}_n(z)\equiv
\psi^{-}_{k-n}(z)$  and $\psi^{+}_0(z)$ the identity  operator.
We have here introduced and made use of
only one pair of deformed parafermionic fields
(besides the identity operator) that corresponds to the pair
$\psi^{\pm}(z)\equiv \psi^{\pm}_1(z)$.
It would be of interest  to identify the appropriate
q-analogues of the other $\psi^{\pm}_n(z)$ and to study
the quantum deformation of the full parafermionic algebra that
these deformed operators would generate.
%%%%%%%%%%%%%%%%%%%%%%%%%%%
\section*{Acknowledgements}
A.H.B. is grateful to NSERC for providing him
with a postdoctoral fellowship. The work of L.V.
is supported through funds provided by
NSERC (Canada) and FCAR (Qu\'ebec).
%%%%%%%%%%%%%%%%%%%%%%%%%%%%%%%%%%%%%%%%%%
%%%%%%%%%%%%%%%%%%%%%%%%%%%%%%%%%%%%%%%%
\pagebreak

%%%%%%%%%%%%%%%%%%%%%%%%%%%%%%%%%%%%%%%%
%%%%%%%%%%%%%%%%%%%%%%%%%%%%%%%%%%%%%%%%
\end{document}